\documentclass[aip,reprint]{revtex4-1}

\usepackage{graphicx}
\usepackage{upgreek}
\usepackage{romannum}
\usepackage{siunitx}
\usepackage[UKenglish]{babel}
\newcommand{\mrm}[1]{\mathrm{#1}}
\newcommand{\micron}[0]{$\upmu$m}
\begin{document}

\title{On-chip polarization rotator for type I second harmonic generation}

\author{Eric~J.~Stanton}
\email[]{eric.stanton@nist.gov}
\affiliation{Applied Physics Division, National Institute of Standards and Technology, Boulder, CO 80305, USA\looseness=-1}

\author{Lin~Chang}
\author{Weiqiang~Xie}
\author{Aditya Malik}
\author{Jon Peters}
\affiliation{Department of Electrical and Computer Engineering, University of California, Santa Barbara, CA 93106, USA\looseness=-1}

\author{Jeff~Chiles}
\author{Nima~Nader}
\affiliation{Applied Physics Division, National Institute of Standards and Technology, Boulder, CO 80305, USA\looseness=-1}

\author{Gabriele~Navickaite}
\author{Davide~Sacchetto}
\author{Michael~Zervas}
\affiliation{LIGENTEC SA, EPFL Innovation Park, B{\^{a}}timent L, 1024 Ecublens, Switzerland}

\author{Kartik~Srinivasan}
\affiliation{Microsystems and Nanotechnology Division, National Institute of Standards and Technology, Gaithersburg, MD, 20899, USA\looseness=-1}

\author{John~E.~Bowers}
\affiliation{Department of Electrical and Computer Engineering, University of California, Santa Barbara, CA 93106, USA\looseness=-1}

\author{Scott~B.~Papp}
\affiliation{Time and Frequency Division, National Institute of Standards and Technology, Boulder, CO 80305, USA\looseness=-1}

\author{Sae~Woo~Nam}
\author{Richard~P.~Mirin}
\affiliation{Applied Physics Division, National Institute of Standards and Technology, Boulder, CO 80305, USA\looseness=-1}
\date{\today}

\begin{abstract}
We demonstrate a polarization rotator integrated at the output of a GaAs waveguide producing type I second harmonic generation (SHG). Form-birefringent phase matching between the pump fundamental transverse electric (TE) mode near 2.0\,\micron{} wavelength and the signal fundamental transverse magnetic (TM) mode efficiently generates light at 1.0\,\micron{} wavelength. A SiN waveguide layer is integrated with the SHG device to form a multi-functional photonic integrated circuit. The polarization rotator couples light between the two layers and rotates the polarization from TM to TE or from TE to TM. With a TE-polarized 2.0\,\micron{} pump, type I SHG is demonstrated with the signal rotated to TE polarization. Passive transmission near 1.0\,\micron{} wavelength shows $\sim$80\,\% polarization rotation across a broad bandwidth of $\sim$100\,nm. By rotating the signal polarization to match that of the pump, this SHG device demonstrates a critical component of an integrated self-referenced octave-spanning frequency comb. This device is expected to provide crucial functionality as part of a fully integrated optical frequency synthesizer with resolution of less than one part in 10$^{14}$.
\end{abstract}

\maketitle

\section{Introduction}

Second harmonic generation (SHG) is an important nonlinear process for stabilizing the carrier-envelope offset frequency of an octave-spanning frequency comb. This process, known as $f$-$2f$ self-referencing, produces a stable frequency reference that finds various uses for timing \cite{Nicklaus2017,Francis2018} as well as for frequency synthesis. Integrated versions of $f$-$2f$ self-referencing are currently under development and are expected to open new applications for this technology \cite{Spencer2018,Newman2019,Malinowski2019}. Devices that produce efficient SHG are generally also suitable for efficient parametric down-conversion (PDC), an enabling process for integrated quantum photonic systems \cite{Guo2017,Stzeni2018,Singh2019}.

A common approach for $f$-$2f$ self-referencing is to compare the 2.0\,\micron{} and 1.0\,\micron{} parts of a frequency comb using SHG \cite{Volet2018,Suh2019,Briles2018,Gaeta2019}. For this technique, octave spanning dissipative Kerr solitons are readily synthesized using a 1.55\,\micron{} pump \cite{Volet2018,Huang2019}.

High conversion efficiency SHG has recently been demonstrated in GaAs \cite{Chang2018,Chang2019,Stanton2019} and AlGaAs \cite{May2019} waveguides using form birefringence phase matching \cite{Fiore1998-1,Fiore1998-2} and in periodically-poled LiNbO$_3$ \cite{Wang2018,Desiatov2019} waveguides using quasi-phase matching. Additionally, AlGaAs platforms \cite{Pu2016,Liu2019,Zheng2018,Chiles2019} have demonstrated frequency combs and supercontinuum generation with high efficiency. Type I SHG in either GaAs or AlGaAs is a promising platform for efficient SHG from 2\,\micron{} to 1\,\micron{} because of the strong $\chi^{(2)}$ nonlinearity \cite{Skauli2002} and small mode sizes. Recently, we demonstrated SHG conversion efficiency of 130\,W$^{-1}$cm$^{-2}$ in GaAs waveguides \cite{Chang2018}. However, this process results in a signal and pump having orthogonal polarization, preventing self-referencing. The octave-spanning comb is typically generated in the fundamental transverse electric (TE) mode, and the SHG process is also most efficient with a fundamental TE mode pump. Therefore, a polarization rotator is needed to rotate the SHG signal from transverse magnetic (TM) to TE.

An exemplary polarization rotator has minimal loss and supports a broad bandwidth with tolerance to fabrication error. This can be achieved with a passive device. Coupling between TE and TM polarizations can be either resonant or adiabatic, with resonant coupling producing a much smaller device and adiabatic coupling supporting a larger bandwidth \cite{Wang2008,Dai2011}. In both cases, the mode coupling between polarizations can be realized by introducing an asymmetry in the waveguide cross-section, either horizontally, vertically, or both. Vertical asymmetry can be engineered by creating an upper and lower cladding with different materials. Horizontal asymmetry is typically formed by partially etching the waveguide core \cite{Wang2008} or by bending the waveguide \cite{Obayya2001}. Using both horizontal and vertical asymmetry is advantageous for adiabatic polarization rotation because the waveguides can be designed with a stronger mode hybridization between the fundamental TE and TM modes and thus support a shorter rotator length. Polarization rotation can also be designed by coupling the fundamental TM mode to the first higher order TE mode and then separately coupling the first order TE mode to the fundamental TE mode \cite{Dai2011,Sacher2014}. These devices are robust to fabrication error and support efficient polarization rotation, but a single stage of polarization rotation between the fundamental TE and TM modes is preferred to reduce the size and complexity of the device.

Here we report a type I SHG waveguide integrated with a passive polarization rotator. The design produces a high polarization rotation from the fundamental TM to the fundamental TE mode in a single coupling stage. Two sets of measurements are performed to characterize the device. First, SHG is demonstrated with a TE-polarized 2\,\micron{} pump to produce a TE-polarized signal at the output of the chip. This confirms the polarization rotator efficiently converts the TM-polarized signal to the TE polarization. Second, the bandwidth of our polarization rotator is characterized by transmission measurements at 1\,\micron{}.

\section{Design}

\begin{figure}[tb]
\includegraphics[width=8.5cm]{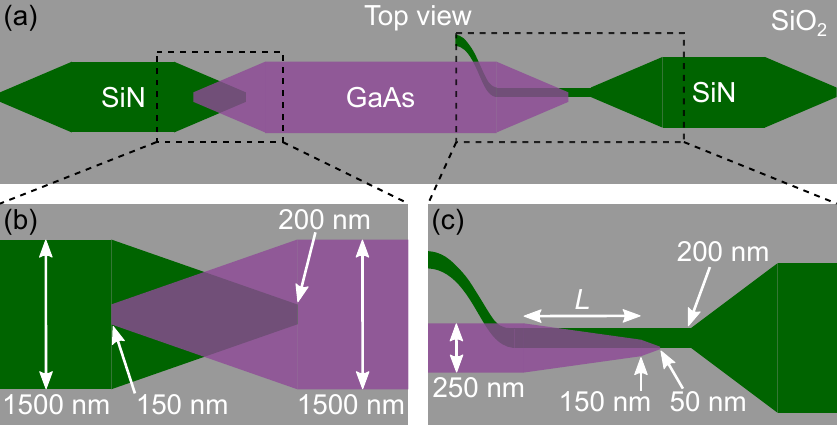}
\caption{(a) Top-view schematic of the SHG device with the polarization rotator. Enlargements of the coupler regions are shown in (b) for the standard coupler and (c) for the polarization rotator. The polarization rotation occurs in (c) between the GaAs widths of 250\,nm and 150\,nm along the length $L$, nominally 800\,\micron{}.}
\label{fig:schematic}
\end{figure}

By integrating both a SiN and a GaAs waveguide, we have broad flexibility in the design to introduce both vertical and horizontal asymmetry. The thickness of the SiN layer is 400\,nm to provide a balance between sufficient confinement at the 2.0\,\micron{} wavelength and efficient coupling between the SiN and GaAs layers at the 1\,\micron{} wavelength. This SiN layer can also support additional photonic elements necessary for $f$-$2f$ self-referencing. The GaAs waveguide is 150\,nm thick and 1500\,nm wide in the straight section, which is designed for phase-matching with a 2.0\,\micron{} pump. The complete polarization rotator schematic is shown in Fig.\,\ref{fig:schematic}(a). It consists of a SiN waveguide at the input, an input SiN-GaAs coupler, a GaAs SHG waveguide, an output GaAs-SiN coupler, and an output SiN waveguide.

The input coupler dimensions are shown in Fig.\,\ref{fig:schematic}(b). Along the length, the cross-section is horizontally symmetric and the GaAs waveguide tapers from 150\,nm to 1500\,nm while the SiN waveguide tapers from 1500\,nm to 200\,nm. An eigenmode expansion calculation estimates high transmission of $>$95\,\% at 2\,\micron{} wavelength from the fundamental TE mode of the SiN waveguide to the fundamental TE mode of the GaAs waveguide. A reference device is designed with the inverse of this input coupler at its output, instead of the polarization rotator, coupling the fundamental TM mode of the 1\,\micron{} signal from the GaAs waveguide to the fundamental TM mode of the SiN waveguide. Transmission of $>$80\,\% is calculated for this output taper.

The output GaAs-SiN coupler designed for polarization rotation is shown in Fig.\,\ref{fig:schematic}(c). Rotation occurs in an 800\,\micron{} long region of the device where the GaAs tapers in width from 250\,nm to 150\,nm and the SiN waveguide width is fixed at 400\,nm. Throughout the taper, the center of the GaAs waveguide is located at one edge of the SiN waveguide to create the horizontal asymmetry. Prior to this region, the SiN waveguide is gradually introduced to the 250\,nm wide GaAs waveguide using an s-bend. After the rotator, the GaAs waveguide is terminated by tapering the width down to a 50\,nm tip over a 20\,\micron{} length.

\begin{figure}[tb]
\includegraphics[width=8.5cm]{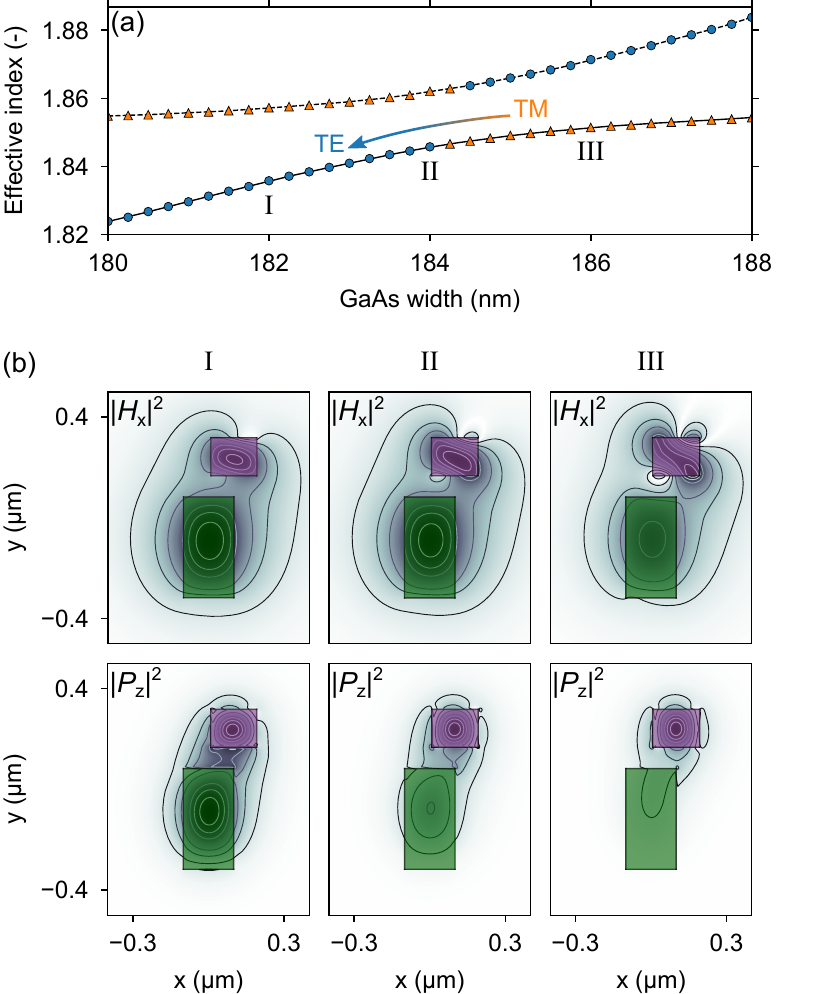}
\caption{(a) Effective index of the hybrid GaAs/SiN waveguide for a varying width of the GaAs layer. The SiN layer has a fixed width of 200\,nm. (b) Mode profiles at three positions (\Romannum{1}, \Romannum{2}, and \Romannum{3}) along the GaAs taper, corresponding to the GaAs widths of 182\,nm, 184\,nm, and 186\,nm, respectively. The lower green rectangle is the SiN waveguide, the upper purple rectangle is the GaAs waveguide, and the background is the SiO$_2$ cladding.}
\label{fig:modes}
\end{figure}

With an optimized geometry, we find a condition where the fundamental TE and TM modes hybridize and exhibit strong coupling. This is shown in Fig.\,\ref{fig:modes}(a) where the effective indices of the first two guided modes are plotted against the GaAs waveguide width. The gap between the fundamental mode and the first higher order mode indicates a hybrid TEM mode at a GaAs waveguide width of 184\,nm. Strong coupling between the TE and TM modes is possible by tapering the GaAs width through this point. The simulated indices indicated with blue circles represent a TE polarization fraction of greater than 50\,\% and the orange triangles correspond to less than 50\,\%. This mode polarization value is defined as $\int{\left| E_x \right|^2 \,\mrm{d}x\,\mrm{d}y} / \int{ ( \left| E_x \right|^2 + \left| E_y \right|^2 ) \,\mrm{d}x\,\mrm{d}y}$. The mode profiles shown in Fig.\,\ref{fig:modes}(b) correspond to the GaAs widths from Fig.\,\ref{fig:modes}(a) as indicated by the Roman numerals. The x-component of the magnetic field shows how the polarization rotates and the z-component of the power shows how light is vertically coupling from the GaAs to the SiN as the GaAs width decreases.

\begin{figure}[tb]
\includegraphics[width=8.5cm]{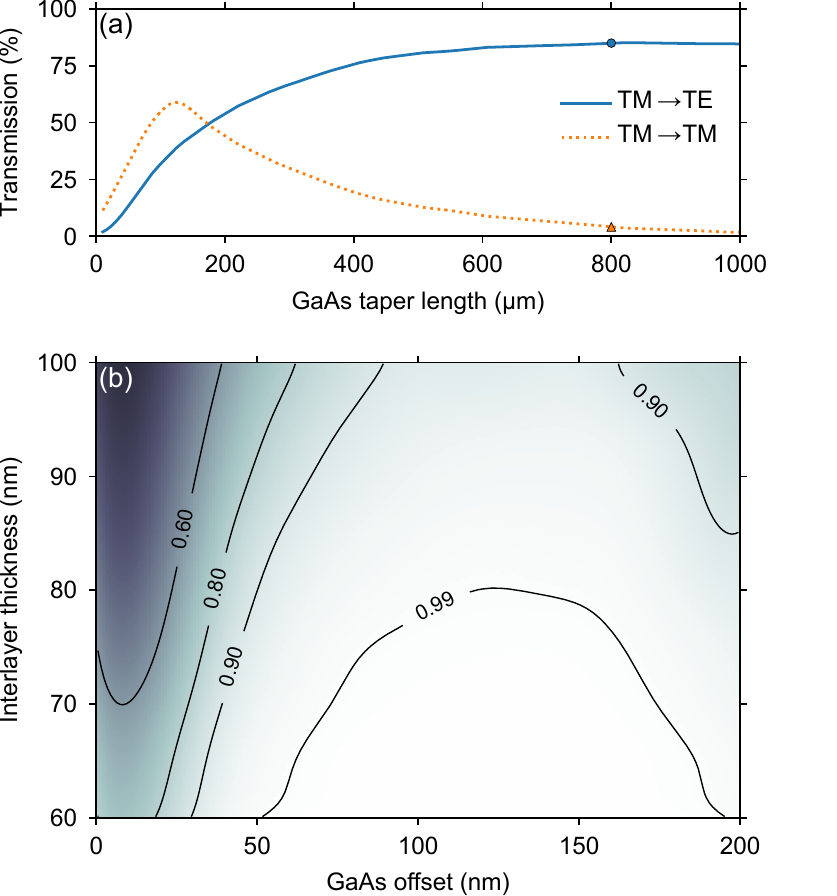}
\caption{(a) Simulated transmission of the polarization rotator to the TE and TM modes of the SiN waveguide for a TE input to the GaAs waveguide as the length $L$ of the GaAs taper is varied. (b) For a taper length of 800\,\micron{}, the polarization rotation is simulated for ranges of spacer layer thickness and GaAs offset, which is the position of the center of the GaAs waveguide relative to the center of the SiN waveguide.}
\label{fig:prop}
\end{figure}

For a GaAs width taper from 250\,nm to 150\,nm, the transmission is simulated from an input TM mode in Fig.\,\ref{fig:prop}(a) to the output TE and TM modes for taper lengths between 0\,\micron{} and 1000\,\micron{}. Efficient polarization rotation is found for a length of 800\,\micron{}, which is the nominal value used for the rotator length. One of the advantages of our design approach is its tolerance to fabrication errors. In particular, two parameters are expected to have relatively large fabrication error: the alignment between the GaAs and SiN waveguide layers and the SiO$_2$ spacer thickness between the GaAs and SiN waveguide layers. The polarization rotation, defined as $T_\mrm{TE}/(T_\mrm{TE} + T_\mrm{TM})$ for a TM input where $T_\mrm{TE/TM}$ is the ratio of the TE/TM polarized output power to the input power, is simulated for a layer misalignment in the range of $\pm$100\,nm and a spacer thickness error in the range of $\pm$25\,\% of the design thickness of 80\,nm, shown in Fig.\,\ref{fig:prop}(b). For a spacer layer thickness less than 80\,nm and a misalignment less than 50\,nm, efficient rotation of at least 90\,\% is achieved.

The width of the GaAs waveguide where the optical modes hybridize and polarization rotation occurs is referred to as the polarization transition width for this adiabatic rotator. This width may shift if the GaAs and SiN waveguides are offset from the nominal design, which occurs from lithography misalignment. Additionally, it can shift if the waveguide widths or heights are different from the design. However, this effect is mitigated by tapering the GaAs width over a larger range than necessary. For example, the simulation shows that the polarization is rotated efficiently when the GaAs width tapers over a range of 10\,nm around the polarization transition width. If the GaAs waveguide is tapered over a 100\,nm range around the same polarization transition width, then the polarization can still be rotated efficiently when accounting for the expected tolerances in our fabrication process. These include an alignment accuracy between layers of less than 70\,nm, transverse to the direction of propagation, a thickness accuracy within 20\,\% of the SiO$_2$ spacer layer, a thickness accuracy within 2\,\% of the SiN and GaAs layers, and a width accuracy within 20\,\% of GaAs and SiN waveguide widths.

\section{Fabrication and experimental setups}\label{sec:fab}

\begin{figure}[tb]
\includegraphics[width=8.5cm]{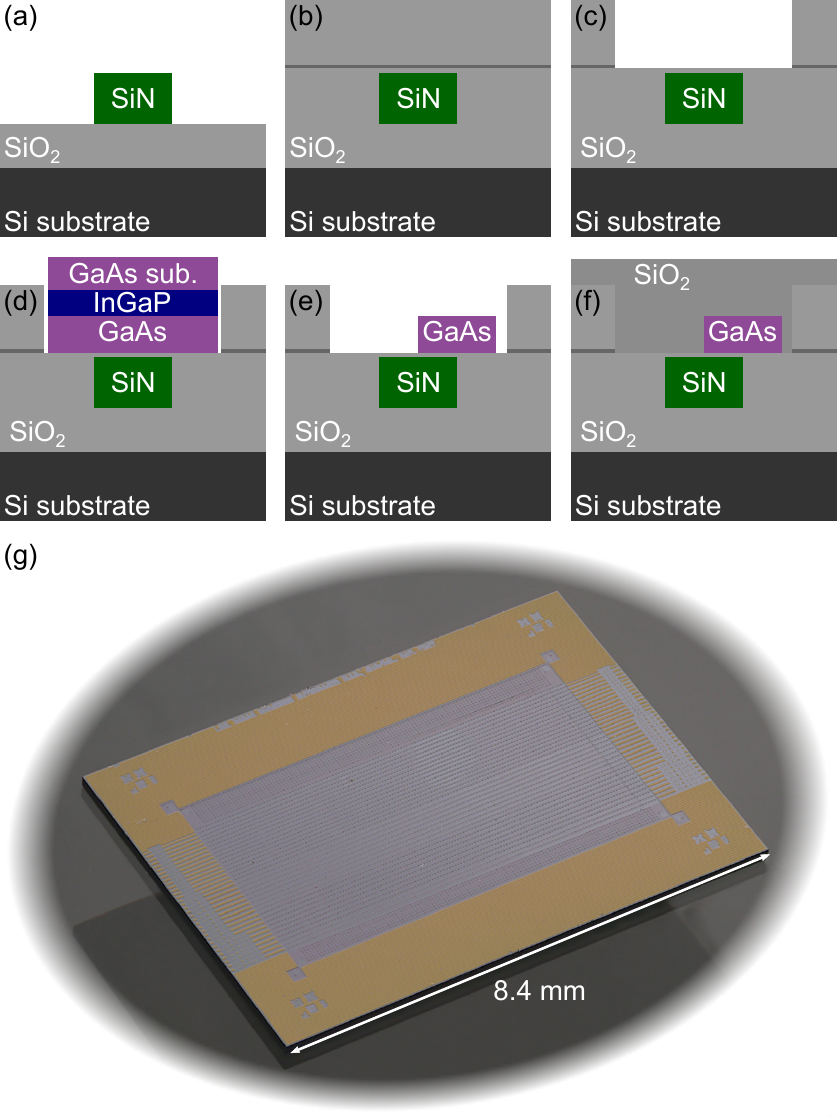}
\caption{Fabrication process flow: (a) SiN is deposited on thermally oxidized Si and etched; (b) SiO$_2$ is deposited and planarized, then an etch-stop and a SiO$_2$ cladding are deposited; (c) a window is etched in the SiO$_2$, leaving a thin SiO2$_2$ spacer above the SiN waveguide layer; (d) a GaAs chip is bonded in the exposed window; (e) the GaAs substrate is removed and the GaAs waveguide is etched; (f) SiO$_2$ is deposited for the top cladding. (g) A photograph of the fabricated chip.}
\label{fig:fab}
\end{figure}

The devices are fabricated on Si wafers using {LIGENTEC} SiN PIC technology, outlined in Fig.\,\ref{fig:fab}. A SiN layer is deposited using low-pressure chemical vapor deposition (LPCVD) on top of a 4\,\micron{} thick thermally oxidized Si layer. The SiN is patterned using deep-UV stepper photolithography to form waveguides using dry etching techniques. An SiO$_2$ spacer layer is deposited with LPCVD and planarized using chemical mechanical polishing, leaving an 80\,nm SiO$_2$ spacer on top of the SiN waveguides. An etch-stop layer is used to preserve the thickness of the spacer layer in subsequent operations. A 2.7\,\micron{} thick SiO$_2$ layer is deposited forming the SiN top cladding. A window is etched through the top cladding to form the bonding area for the GaAs 80\,nm above the SiN waveguides.

The GaAs waveguide layer is formed by directly bonding a GaAs chip to the SiO$_2$ spacer layer. First, $\sim$5\,nm of Al$_2$O$_3$ is deposited on the GaAs surface by atomic layer deposition to improve the bonding energy \cite{Sahoo2018,Zheng2018}. Then, both bonding surfaces are cleaned and O$_2$ plasma activated before bonding in atmosphere at room temperature. An anneal is performed at \SI{100}{\celsius} under pressure from a graphite fixture to strengthen the bond. The GaAs substrate is removed in two steps: first with mechanical lapping until $\sim$50\,\micron{} remains, then a mixture of H$_2$O$_2$/NH$_4$OH/H$_2$O removes the remaining substrate, stopping on the InGaP layer. Previous work has used an Al$_{0.8}$Ga$_{0.2}$As layer for the etch-stop, which is removed with dilute HF \cite{Chang2018,Chang2019}. However, the InGaP etch-stop layer used in this work is instead removed with HCl, which is more compatible with the current process since HCl does not etch the exposed SiO$_2$. Next, the GaAs waveguide is patterned and etched with electron-beam lithography and ICP-RIE. A final PECVD SiO$_2$ forms the top cladding. A metal layer is also fabricated on top of the SiO$_2$ top cladding as part of this foundry process, but the metal layer is not used for this device.

\begin{figure}[tb]
\includegraphics[width=8.5cm]{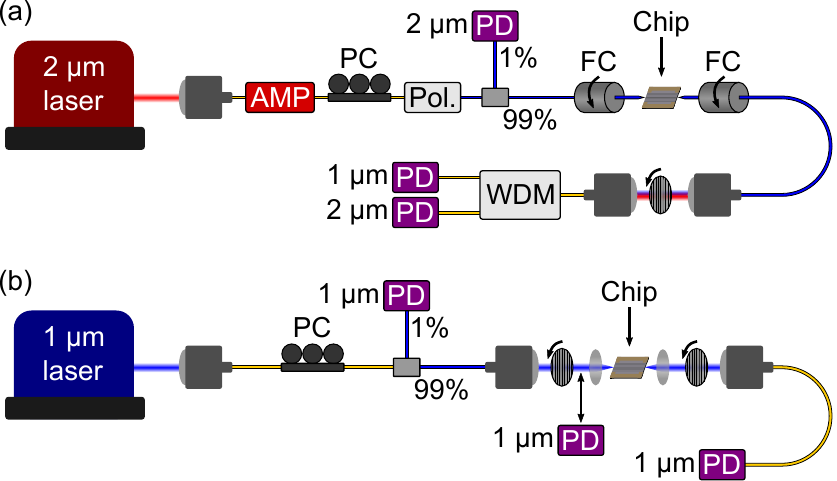}
\caption{Experimental setups for (a) the SHG demonstration and (b) the passive polarization measurement. Abbreviations: AMP is a fiber amplifier; PC is a polarization controller; Pol. is a fiber-based linear polarizer; PD stands for photodetector; FC is for fiber coupler; and WDM is a wavelength division multiplexer for splitting the 1\,\micron{} and 2\,\micron{} light. The yellow connections are single mode fibers and the blue connections are polarization maintaining single mode fibers.}
\label{fig:setups}
\end{figure}

For the SHG experimental setup, shown in Fig.\,\ref{fig:setups}(a), a tunable 2\,\micron{} wavelength laser is amplified. The polarization is adjusted to couple light to the slow-axis of the polarization maintaining (PM) fiber. A splitter taps $\sim$1\,\% of the input light to a photodetector to monitor the input power. Lensed PM fibers couple light into and out of the waveguides with the slow-axis aligned to the TE polarization. After the PM fiber, a free-space nanoparticle film polarizer is used to distinguish between TE and TM polarized light at the waveguide output. This polarizer has a transmission of $>$83\,\% and a polarization extinction ratio of $>$10$^3$ for both the 1\,\micron{} and 2\,\micron{} wavelengths. A broadband wavelength division multiplexer is used to separate the 1\,\micron{} signal from the 2\,\micron{} pump so both outputs can be monitored while the input wavelength is scanned.

Passive transmission measurements are performed using the setup shown in Fig.\,\ref{fig:setups}(b). Two different 1\,\micron{} tunable lasers are used to cover wavelengths of 910--980\,nm and 1045--1083\,nm. The laser polarizations are aligned to the TM polarization of the waveguides. The input power is monitored on a photodetector from a $\sim$1\,\% power tap. Light is coupled on and off the chip in free-space using aspheric lenses and linear polarizers. The output light is fiber coupled and the power is monitored with a photodetector. Across a $\sim$100\,nm bandwidth centered at 1\,\micron{}, the polarization extinction ratio of the measurement setup is $>$10$^3$.

\section{Experiment}

\begin{figure}[tb]
\includegraphics[width=8.5cm]{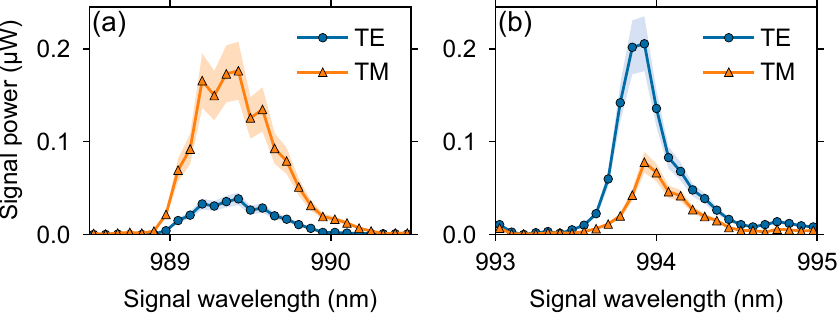}
\caption{SHG signals from a TE polarized pump of (a) the reference device with a symmetric output coupler and (b) the polarization rotator device.}
\label{fig:SHG}
\end{figure}

SHG is demonstrated in both the reference device and the polarization rotator. The spectrum of the generated signal for each is shown in Fig.\,\ref{fig:SHG}. After the signal is coupled to the output fiber, it passes through a linear polarizer aligned to the on-chip TE  mode or TM mode to distinguish the power generated from each mode. The standard error of the signal power in each polarization is indicated with the shaded regions, as determined by polarization calibration measurements. Due to slight differences in the width of each waveguide (on the order of 5\,nm), the phase matched pump wavelength of the reference device is 9\,nm shorter than the polarization rotator device. At the peak of the reference signal, the light is $82.1 \pm 4.8$\,\% TM and $17.9 \pm 4.8$\,\% TE. As predicted, the polarization rotator device shows the opposite trend with an SHG signal at the peak of $27.5 \pm 5.3$\,\% TM and $72.5 \pm 5.3$\,\% TE light.

The SHG signal of the reference device in Fig.\,\ref{fig:SHG}(a) has a full-width at half-max (FWHM) of 0.6\,nm, which is wider than the FWHM from the polarization rotator device of 0.4\,nm. This difference is not expected to be due to the different output couplers, but it is instead likely due to variations of the SHG waveguide width, thickness, or both. Similar observations have been made in our previous SHG experiments \cite{Chang2018}. Notice that the reference device has a lower peak conversion efficiency due to greater index variations compared to the polarization rotator device. This also indicates that the insertion loss of the polarization rotator is similar to the insertion loss of the reference device.

Besides the efficient operation of the polarization rotator, both devices have lower SHG conversion efficiency than expected: 0.11\,W$^{-1}$ in this work compared to 2.50\,W$^{-1}$ from our previous work \cite{Chang2018}. From transmission measurements, we found that the waveguide propagation loss is similar, on the order of 20\,dB/cm near 1\,\micron{} (due to surface-state absorption) and 0.5\,dB/cm near 2\,\micron{}. However, as we have noticed previously, there is an optimal length to achieve the maximum conversion efficiency. For short waveguides $<$2\,mm, the conversion efficiency scales with the length squared, as expected from theoretical calculations. However, other mechanisms begin to degrade the efficiency as the length increases further, which we expect is due to propagation loss and variations in the effective indices of both the pump and signal modes due to geometric variations. These variations may be due to short-scale roughness at the interfaces between the waveguide core and cladding, but also larger-scale variations in the width and height of the waveguide due to lithography errors or an uneven termination after removing the etch-stop. Further investigation is required to fully understand the origin and magnitude of this degradation effect. One likely cause, as mentioned in Section\,\ref{sec:fab}, is that a new etch-stop composed of InGaP was used in this work compared to the AlGaAs used in our previous work. The roughness on the top surface of the GaAs waveguide is 5.6\,nm RMS, which contributes to phase decoherence. This roughness is likely inherent to the epitaxial wafer and we expect that the same etch-stop removal process will produce a more smooth waveguide surface when a new epitaxial wafer is used with a smooth interface between the GaAs waveguide layer and the InGaP etch-stop layer.

\begin{figure}[tb]
\includegraphics[width=8.5cm]{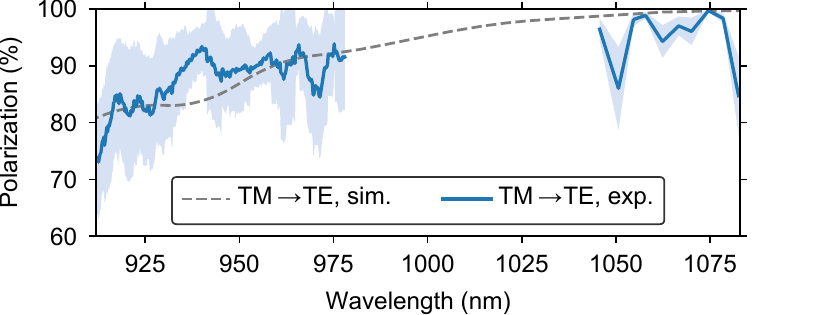}
\caption{Experimental and simulated wavelength dependence of the polarization rotation for a TM input and a TE output.}
\label{fig:passive}
\end{figure}

To understand the dependence of the polarization rotation on the signal wavelength, passive transmission measurements are performed with two different tunable lasers with wavelengths near 1\,\micron{}. The polarization at the output of the polarization rotator is defined as $T_\mrm{TE}/(T_\mrm{TE} + T_\mrm{TM})$. Light is input to the chip with a TM polarization and both TE and TM polarizations are measured separately at the output. These results are shown in Fig.\,\ref{fig:passive} along with a grey dashed line indicating the simulated polarization trend. The measured polarization agrees well with the simulation, though small fluctuations on the order of $\pm$10\,\% are present due to Fabry-Perot resonances formed by reflections at the taper tips of the input SiN-GaAs coupler. While the simulation does not account for these reflections, other small fluctuations are also present in the simulated polarization trend that are not due to Fabry-Perot resonances. The physical effect that produces these fluctuations is the partial resonant coupling of the rotator. For wavelengths longer than 1075\,nm, the coupling is accurately assumed to be purely adiabatic because the mode splitting, observed in Fig.\,\ref{fig:modes}(a), is sufficiently large. For the shorter wavelengths, the mode splitting is less significant, so a longer coupler would be necessary to achieve a purely adiabatic transition. The designed geometry achieves the largest mode-splitting and therefore most adiabatic transition permitted by other fabrication constraints.

\section{Conclusions}

We have demonstrated a polarization rotator in a hybrid GaAs/SiN waveguide to couple light from a GaAs SHG waveguide to a SiN waveguide. The platform allows for the polarization rotator to have both vertical and horizontal asymmetry, which supports coupling between the fundamental TE and TM modes. SHG is demonstrated in this type I waveguide with a TE pump and a final TE signal at the output of the polarization rotator. The InGaP etch-stop process likely limits the conversion efficiency, and future work will minimize the roughness on the top surface of the GaAs waveguide. Polarization rotation is measured over a broad bandwidth, close to 100\,nm, with excellent agreement to the design simulation. This device is suitable for integration with a SiN-based photonic integrated circuit to route and detect light for $f$-$2f$ self-referencing of an octave-spanning frequency comb.

\begin{acknowledgments}
This work was funded by the DARPA MTO DODOS program. We thank Travis~M.~Autry, David~R.~Carlson, and Micheal~Geiselmann for useful discussions and inputs on the manuscript.
\end{acknowledgments}
\bibliography{EJS_biblio}
\end{document}